%% file: eprint.tex
\documentclass[12pt]{article}
\usepackage{graphicx}

\usepackage{ifthen} % for conditional statements
\newboolean{pdflatex}
\setboolean{pdflatex}{true} % False for eps figures 

\newboolean{articletitles}
\setboolean{articletitles}{true} % False removes titles in references

\newboolean{uprightparticles}
\setboolean{uprightparticles}{false} %True for upright particle symbols

\newboolean{inbibliography}
\setboolean{inbibliography}{false} %True once you enter the bibliography

\input{preamble}
\newcommand\pubnumber{SNSN-323-63}
\newcommand\pubdate{\today}
\def\heidelberg{Ruprecht-Karls-Universitaet Heidelberg\\ Physikalisches Institut\\
Im Neuenheimer Feld 226 \\69120 Heidelberg, Germany
}

\def\Title#1{\begin{center} {\Large #1 } \end{center}}
\def\Author#1{\begin{center}{ \sc #1} \end{center}}
\def\Address#1{\begin{center}{ \it #1} \end{center}}

\newcommand\pubblock{\rightline{\begin{tabular}{l} \pubnumber\\
         \pubdate  \end{tabular}}}
\newenvironment{Abstract}{\begin{quotation}  }{\end{quotation}}
\newenvironment{Presented}{\begin{quotation} \begin{center} 
             PRESENTED AT\end{center}\bigskip 
      \begin{center}\begin{large}}{\end{large}\end{center} \end{quotation}}

\input econfmacros.tex
\begin{document}
\begin{titlepage}
\pubblock

\vfill
\Title{Direct CPV in two-body
and multi-body charm
decays at LHCb}
\vfill
\Author{ Evelina Gersabeck on behalf of the LHCb collaboration}
\Address{\heidelberg}
\vfill
\begin{Abstract}

The Standard Model predicts CP asymmetries in charm decays of $O(10^{-3})$  and the observation of significantly larger CP violation could indicate non-Standard Model physics effects. 
During 2011 and 2012, the LHCb experiment collected a sample corresponding to $3/fb
$  yielding the worldÕs largest sample of decays of charmed hadrons. This allowed the 
CP violation in charm to be studied with unprecedented precision in many two- body and multibody decay modes.  
The most recent LHCb searches for direct CP violation are presented in these proceedings.
\end{Abstract}
\vfill
\begin{Presented}
Presented at the 8th International Workshop on the CKM Unitarity Triangle (CKM 2014), Vienna, Austria, \\September 8-12, 2014
\end{Presented}
\vfill
\end{titlepage}
\def\thefootnote{\fnsymbol{footnote}}
\setcounter{footnote}{0}

\section{Introduction}

The excellent performance of the LHC and the LHCb experiment, along with large production $c\bar{c}$ cross sections for $pp$ collisions at $\sqrt{s}$ of 7 and 8 TeV has enabled unprecedentedly large samples of charm decays to be recorded during 2011 and 2012, corresponding to $3/fb$ of integrated luminosity. These large samples allow the study  of CP violation (CPV) effects at a precision not achieved before in charm decays.
% and thus allowing to measure even small CPV (CPV) effects in charm decays. 
The data was taken with a regular swap of the polarity of the spectrometer dipole magnet which can  compensate for the left-right detector asymmetries to a first order. Both charm decays, promptly produced in the primary pp interaction, and coming from a parent beauty hadron are exploited at LHCb; this is indicated for each of the presented analyses.

%To separate prompt and secondary charm, the impact parameter is used. Direct charm points back to the primary vertex which corresponds to a very small impact parapeter. Secondary charm origins from a B-decay which has a long lifetime and other decay products apart from the D-meson. This D-meson doesn't necessarily point back to the primary vertex and its impact parameter is non-zero.

%\section{Direct CPV}
\section{Time-integrated CP asymmetry in $\Dz\to h^+h^-$ from semileptonic decays}
A search for a time-integrated CP asymmetry in $\Dz\to h^+h^-$ decays is performed using the full dataset  corresponding to $3/fb$. The flavour of the initial $\Dz$ state is tagged by the charge of the muon in the semileptonic $B\to\Dz\mu^-\nu_{\mu}X$ decays.

The raw measured asymmetry for tagged $\Dz$ messns to a final state $f$ is given by:
\begin{equation}
A_{raw}(f)=\frac{N(B\to\Dz\mu^-X)-N(B\to\Dzb\mu^+X)}{N(B\to\Dz\mu^-X)+N(B\to\Dzb\mu^+X)},
\end{equation}
where $N$ indicates the number of reconstructed events of a given decay after background subtraction, and $X$ refers to the undetected final state particles from the semileptonic B decay.
The raw asymmetry is a sum of the physical CP asymmetry, ($A_{CP}(f)$), the production asymmetry ($A_P(B)$) and detection asymmetry ($A_D(\mu)$) :
\begin{equation}
A_{raw}(f)=A_{CP}(f)+A_P(B)+A_D(\mu).
\end{equation}
As the quantity of interest is $A_{CP}(f)$, the main experimental challenge is to separate it from the nuisance asymmetries.
An experimentally more robust variable, $\Delta A_{CP}$ can be constructed by taking the difference of the raw asymmetries measured in $\Dz\to K^+K^-$ and $\Dz\to \pi^+\pi^-$ decays:
\begin{equation}
\Delta A_{CP} = A_{raw}(KK)-A_{raw}(\pi\pi),
\end{equation}
and thus cancelling the production and the muon detection asymmetries to a first order. Alternatively, for extracting $A_{CP}(KK)$, the detection and production asymmetries can be measured using Cabibbo-favoured (CF) $B\to\Dz(\to K^-\pi^+)\mu^-X$ decays where no CPV is expected. An additional detection asymmetry, $A_D(K\pi)$, arises due to the different interaction rates of the charged $K$ with the matter. To remove this asymmetry, the control channels $D^+\to K^-\pi^+\pi^+$ and $D^+\to \bar{K^0}\pi^+$ are used. In the $D^+\to \bar{K^0}\pi^+$ decays, the detection asymmetry of $K^0$ arising due to the combined
effect of CPV in mixing in the neutral kaon system and the different interaction
rates of $K^0$ and $\bar{K^0}$ in the detector material, $A_{CP/int}$, is estimated from simulation and subtracted from the raw asymmetry. Once $\Delta A_{CP}$ and $A_{CP}(KK)$ are measured, the individual asymmetry $A_{CP}(\pi \pi) = \Delta A_{CP} - A_{CP}(KK)$ can be computed.

In total, $\sim 2.1\mathrm \times 10^6 ~ \Dz \to KK$ and  $\sim 0.7 \times 10^6 ~\Dz \to \pi\pi$ decays are reconstructed. The analysis is done separately for the 2011 and the 2012 data, and for the two magnet polarities and consistent results are obtained. As the production and detection asymmetries depend on the kinematic distributions, weights are assigned to the candidates such that the kinematic distributions are equalised. 
The results for the asymmetries~\cite{LHCb-PAPER-2014-013} 
\begin{equation}
\Delta A_{CP} =(+0.14\pm0.16(stat)\pm0.08(syst))\%,
\end{equation}
\begin{equation}
A_{CP}(KK)=(-0.06\pm0.15(stat)\pm0.10(syst))\%,
\end{equation}
and, with correlation rho=0.28,
\begin{equation}
A_{CP}(\pi\pi)=(-0.20\pm0.19(stat)\pm0.10(syst))\%
\end{equation}
are compatible with CP conservation. The $A_{CP}(hh)$ asymmetries are the most precise measurement of individual  asymmetries up to date. The precision of $\Delta A_{CP}$ is comparable to the preliminary result for $\Delta A_{CP}$ measured using prompt $\Dz$ decays reconstructed in 1/fb of integrated luminosity~\cite{LHCb-CONF-2013-003} .

\section{CP asymmetries in $D^+_{(s)}\to K^0_S h^+$ decays}
Following a similar analysis strategy, the raw asymmetries in the prompt $D^+_{(s)}\to K^0_S h^+$ decays
\begin{equation}
A_{raw}(K^0_S h^+) = A_{CP}(K^0_S h^+)+A_P(D^+_{(s)})+A_D(h^+) + A_{CP/int}(K^0/\bar{K^0}),
\end{equation}
where $h^+ = K^+$ or $\pi^+$, are used to extract the quantities of interest
\begin{eqnarray}
A_{CP}^{DD} &=& ((A_{raw}(D^+_s\to K^0_S\pi^+) - (A_{raw}(D^+_s\to K^0_S K^+)) \\ \nonumber
&-& ((A_{raw}(D^+\to K^0_S\pi^+) - (A_{raw}(D^+\to K^0_S K^+))\\ \nonumber
&\approx& A_{CP}(D^+_s\to K^0_S\pi^+) + A_{CP}(D^+\to K^0_S K^+),
\end{eqnarray}
%which to a first order is (up to a small contribution due to the CPV, mixing and interaction with the material of the $K^0$):
%\begin{equation}
%A_{CP}^{DD}=A_{CP}(D^+_s\to K^0_S\pi^+) + A_{CP}(D^+\to K^0_S K^+),
%\end{equation}
and
\begin{equation}
A_{CP}(D^+_S\to K^0_S\pi^+) = A_{raw}(D^+_s\to K^0_S\pi^+)-A_{raw}(D^+_s\to \Phi \pi^+),
\end{equation}
\begin{eqnarray}
A_{CP}(D^+\to K^0_S K^+) &=& (A_{raw}(D^+\to K^0_S K^+)-A_{raw}(D^+_s \to K^0_S K^+)) \\ \nonumber
&- &(A_{raw}(D^+\to K^0_S\pi^+) - A_{raw}(D^+_S\to \Phi \pi^+)).
\end{eqnarray}

The analysis is done using the full available data sample of $3/fb$. In total, $\sim4.8\times 10^6 ~ D^+ \to K_S^0\pi^+$,  $\sim0.12 \times 10^6 ~D^+_S \to K_S^0\pi^+$,  $\sim1.0 \times 10^6 ~ D^+ \to K_S^0 K^+$,  $\sim1.5 \times 10^6 ~D^+_S \to K_S^0 K^+$, $\sim7\times 10^6 ~ D^+ \to \Phi \pi^+$,  and $\sim13. \times 10^6 ~D^+_S \to \Phi \pi^+$ decays were reconstructed. The analysis is done separately for 2011 and 2012 data, and for both magnet polarities and consistent results are obtained.. %A kinematical re-weighting was employed in order to equalise kinematical distributions of the signal and the control channels. 
The results 
\begin{equation}
A_{CP}(D^+\to K^0_S K^+) = (+0.03\pm0.17(stat)\pm0.14(syst))\%,
\end{equation}
\begin{equation}
A_{CP}(D^+_S\to K^0_S\pi^+) = (+0.38\pm0.46(stat)\pm0.17(syst))\%,
\end{equation}
\begin{equation}
A_{CP}(D^+\to K^0_S K^+) + A_{CP}(D^+_S\to K^0_S\pi^+) = (0.41\pm0.49(stat)\pm0.26(syst))\%.
\end{equation}
show no indication of CPV~\cite{LHCb-PAPER-2014-018}. These are the most precise measurements of these quantities.

%%%%%%%%%%%%%%%%%%%%%%%%%%%%%%%%%%%%%%%%%%%%%%%%%%%%%%%%%%%%%%%%%%%%%%%%%
%%
%%   use this format to include an .eps figure into your paper
%%
%\begin{figure}[htb]
%\centering
%\includegraphics[height=1.5in]{magnet}
%\caption{Plan of the magnet used in the mesmeric studies.}
%\label{fig:magnet}
%\end{figure}
%%%%%%%%%%%%%%%%%%%%%%%%%%%%%%%%%%%%%%%%%%%%%%%%%%%%%%%%%%%%%%%%%%%%%%%%%%%

%%%%%%%%%%%%%%%%%%%%%%%%%%%%%%%%%%%%%%%%%%%%%%%%%%%%%%%%%%%%%%%%%%%%%%%%%
%%
%%   use this format to include a LaTeX table  into your paper
%%
%\begin{table}[t]
%\begin{center}
%\begin{tabular}{l|ccc}  
%Patient &  Initial level($\mu$g/cc) &  w. Magnet &  
%w. Magnet and Sound \\ \hline
 %Guglielmo B.  &   0.12     &     0.10      &     0.001  \\
 %Ferrando di N. &  0.15     &     0.11      &  $< 0.0005$ \\ \hline
%\end{tabular}
%\caption{Blood cyanide levels for the two patients.}
%\label{tab:blood}
%\end{center}
%\end{table}
%%%%%%%%%%%%%%%%%%%%%%%%%%%%%%%%%%%%%%%%%%%%%%%%%%%%%%%%%%%%%%%%%%%%%%%%%%%

\section{Search for CPV in $\Dz \to \pi^-\pi^+\pi^0$ decays with the energy test}

The energy test~\cite{energytest} is an unbinned model-independent statistical method to search for time- integrated CP violation in $\Dz \to \pi^-\pi^+\pi^0$ decays. The method relies on the comparison of two $\Dz$ and $\bar{\Dz}$ flavour samples and is sensitive to CPV localised in the phase-space of the multi body final state. The flavour of the prompt $\Dz$ is tagged by the charge of the slow pion in the decay $D^* \to \Dz \pi_s$. For the reconstruction of the $\Dz$ both merged and resolved neutral pions are used.
%The decay $\Dz \to \pi^-\pi^+\pi^0$ proceeds via a singly Cabibbo-suppressed $c \to du\bar{d}$ transition. 
The previous most sensitive study of this decay has been done by the BaBar collaboration~\cite{babar}. At LHCb, the energy test is used to assign a p-value for a non-zero CPV hypothesis~\cite{LHCB-PAPER-2014-054}. In this method, a test statistic $T$ is used to compare the average distances based on the metric function $\psi$.
It is defined as
\begin{equation}
T = \sum_{i,j>i}^{n}\frac{\psi_{ij}}{n(n-1)}
 + \sum_{i,j>i}^{\overline{n}}\frac{\psi_{ij}}{\overline{n}(\overline{n}-1)}
 - \sum_{i,j}^{n,\overline{n}}\frac{\psi_{ij}}{n\overline{n}} ,
\label{eqn:T}
\end{equation}
and the metric function $\psi_{ij}\equiv\psi(d_{ij})=e^{-d_{ij}^2/2\sigma^2}$ is chosen as a Gaussian function with a tunable parameter $\sigma$ as it should be a falling function with increasing the distance between events. $T$ compares the average distances of pairs of events belonging to two samples of opposite flavour. The normalisation factor removes the impact of global asymmetries. The distance between two points in phase space is given by $d_{ij}=(m_{12}^{2,j}-m_{12}^{2,i},m_{23}^{2,j}-m_{23}^{2,i},m_{13}^{2,j}-m_{13}^{2,i})$, 
where the $1,2,3$ subscripts indicate the final-state particles. For no-CPV, $T$ is expected to be zero, and larger than zero in case of the CPV. This unbinned technique calculates a $p$-value under the hypothesis of $CP$ symmetry by comparing the nominal $T$ value observed in data to a distribution of $T$ values obtained from permutation 
samples, where the flavour of the $\Dz$ is randomly reassigned to simulate samples without $CP$ violation. The $p$-value for the no CPV hypothesis is obtained as the fraction of permutation $T$ values greater than the nominal $T$ value. The $p$-value from the fitted $T$ distribution can be calculated as the fraction of the integral 
of the function above the nominal $T$ value. This approach is used for the nominal result. If large $CP$ violation is observed, the observed $T$ value is likely to lie outside 
the range of permutation $T$ values.  In this case the permutation $T$ distribution can be fitted 
with a generalised extreme value function
\begin{eqnarray}
f(T;\mu,\delta,\xi) = N \left[1+\xi\left(\frac{T-\mu}{\delta}\right)\right]^{(-1/\xi)-1}&&\nonumber\\
\times\exp\left\{-\left[1+\xi\left(\frac{T-\mu}{\delta}\right)\right]^{-1/\xi}\right\},&&
\end{eqnarray}
with normalisation $N$, location parameter $\mu$, scale parameter $\delta$, and shape parameter $\xi$. The $p$-value from the fitted $T$ distribution can be calculated as the fraction of the integral 
of the function above the nominal $T$ value. This approach is used for the sensitivity studies. Using 100 permutations, the sensitivity studies are reported in Table~\ref{tab:sensitivity_overview}.

\begin{table}[t]
\caption{Overview of sensitivities to various $CP$ violation scenarios. 
$\Delta A$ and $\Delta \phi$ denote, respectively, 
change in amplitude and phase of the resonance $R$.}
\centering
\begin{tabular}{lll}
\hline \hline 
$R$ ($\Delta A$, $\Delta \phi$) & $p$-value (fit) & Upper limit \\ 
\hline  
$\rho^0$ $(4\%$, $0^\circ)$ & $3.3^{+1.1}_{-3.3}\times10^{-4}$ & $4.6\times10^{-4}$  \\ 
$\rho^0$ $(0\%$, $3^\circ)$ & $1.5^{+1.7}_{-1.4}\times10^{-3}$ & $3.8\times10^{-3}$  \\ 
$\rho^+$ $(2\%$, $0^\circ)$ & $5.0^{+8.8}_{-3.8}\times10^{-6}$ & $1.8\times10^{-5}$  \\ 
$\rho^+$ $(0\%$, $1^\circ)$ & $6.3^{+5.5}_{-3.3}\times10^{-4}$ & $1.4\times10^{-3}$  \\ 
$\rho^-$ $(2\%$, $0^\circ)$ & $2.0^{+1.3}_{-0.9}\times10^{-3}$ & $3.9\times10^{-3}$  \\ 
$\rho^-$ $(0\%$, $1.5^\circ)$ & $8.9^{+22}_{-6.7}\times10^{-7}$ & $4.2\times10^{-6}$  \\ 
\hline \hline  
\end{tabular}
\label{tab:sensitivity_overview}
\end{table}

By counting the fraction of permutations with a $T$ value above the nominal $T$ value in the data, 
a $p$-value of $(2.6\pm0.5)\times10^{-2}$ is extracted. This result is based on 1000 permutations.
The results correspond to a data sample of about 2/fb collected during 2012.
%A small phase-space region dominated by the $\rho^+$ resonance contains candidates with a local positive asymmetry exceeding $1\sigma$ significance. 
Varying the metric parameter 
results in the $p$-values listed in Table~\ref{tab:results}; all the $p$-values are at the $10^{-2}$ level. 

\begin{table}[t]
\caption{Results for various metric parameter values. 
The $p$-values are obtained with the counting method.}
\centering
\begin{tabular}{cc}
\hline \hline 
$\sigma$~[$\mathrm{GeV}^2/\mathrm{c}^4$] & $p$-value \\ 
\hline  
0.2 & $(4.6 \pm 0.6)\times10^{-2}$ \\
0.3 & $(2.6 \pm 0.5)\times10^{-2}$ \\
0.4 & $(1.7 \pm 0.4)\times10^{-2}$ \\
0.5 & $(2.1 \pm 0.5)\times10^{-2}$ \\
\hline \hline  
\end{tabular}
\label{tab:results}
\end{table}

The data sample has been split according to various criteria to test the stability of the results.
Analyses of sub-samples with opposite magnet polarity, with different trigger configurations, and with fiducial 
selection requirements removing areas of high local asymmetry of the tagging soft pion from the $D^*+$ decay 
all provide consistent results. Various checks have been performed to ensure there are no asymmetries arising form background events or detector related asymmetries.

The analysis has the words best sensitivity from a single experiment to local CPV in this decay.

\section{Conclusions}
LHCb has performed world leading precision measurements in the charm sector. The searches for direct CPV in two- and multi body decays are consistent with CP conservation, in agreement with the SM at the current level of precision.

%\Acknowledgements
%I am grateful to Don Alfonso d'Alba for certain services essential to 
%this investigation.

%\section*{References}
\bibliography{main,LHCb-PAPER,LHCb-CONF,LHCb-DP,LHCb-TDR}

%\begin{thebibliography}{99}

%%
%%  bibliographic items can be constructed using the LaTeX format in SPIRES:
%%    see    http://www.slac.stanford.edu/spires/hep/latex.html
%%  SPIRES will also supply the CITATION line information; please include it.
%%

%\bibitem{Mesmer}
%F. A. Mesmer, Proc. Wien. Acad. Sci. {\bf 13}, 1564, 1593 (1762).
%%CITATION = PWASA,13,1564;%%

%\end{thebibliography}

\end{document}

%% file: preamble.tex
\usepackage{microtype}
\usepackage{lineno}  % for line numbering during review
\usepackage{xspace} % To avoid problems with missing or double spaces after
\usepackage{caption} %these three command get the figure and table captions automatically small

\usepackage{graphicx}  % to include figures (can also use other packages)
\usepackage{color}
\usepackage{colortbl}
\graphicspath{{./figs/}} % Make Latex search fig subdir for figures

\usepackage{amsmath} % Adds a large collection of math symbols
\usepackage{amssymb}
\usepackage{amsfonts}
\usepackage{upgreek} % Adds in support for greek letters in roman typeset

\newcommand*\patchAmsMathEnvironmentForLineno[1]{%
\expandafter\let\csname old#1\expandafter\endcsname\csname #1\endcsname
\expandafter\let\csname oldend#1\expandafter\endcsname\csname
end#1\endcsname
 \renewenvironment{#1}%
   {\linenomath\csname old#1\endcsname}%
   {\csname oldend#1\endcsname\endlinenomath}%
}
\newcommand*\patchBothAmsMathEnvironmentsForLineno[1]{%
  \patchAmsMathEnvironmentForLineno{#1}%
  \patchAmsMathEnvironmentForLineno{#1*}%
}
\AtBeginDocument{%
\patchBothAmsMathEnvironmentsForLineno{equation}%
\patchBothAmsMathEnvironmentsForLineno{align}%
\patchBothAmsMathEnvironmentsForLineno{flalign}%
\patchBothAmsMathEnvironmentsForLineno{alignat}%
\patchBothAmsMathEnvironmentsForLineno{gather}%
\patchBothAmsMathEnvironmentsForLineno{multline}%
\patchBothAmsMathEnvironmentsForLineno{eqnarray}%
}

\usepackage{hyperref}    % Hyperlinks in references
\usepackage[all]{hypcap} % Internal hyperlinks to floats.

\input{lhcb-symbols-def} % Add in the predefined LHCb symbols

\usepackage{cite} % Allows for ranges in citations
\usepackage{mciteplus}

%% file: lhcb-symbols-def.tex
\ifthenelse{\boolean{uprightparticles}}%
{

 \def\PDelta      {\ensuremath{\Delta}\xspace}                 
 \def\PXi      {\ensuremath{\Xi}\xspace}                 
 \def\PLambda      {\ensuremath{\Lambda}\xspace}                 
 \def\PSigma      {\ensuremath{\Sigma}\xspace}                 
 \def\POmega      {\ensuremath{\Omega}\xspace}                 
 \def\PUpsilon      {\ensuremath{\Upsilon}\xspace}                 
 
 \def\PB      {\ensuremath{\mathrm{B}}\xspace}                 
                  
 \def\PD      {\ensuremath{\mathrm{D}}\xspace}

 \def\PK      {\ensuremath{\mathrm{K}}\xspace}

 \def\Pi      {\ensuremath{\mathrm{i}}\xspace}

}
{

 \mathchardef\PDelta="7101
 \mathchardef\PXi="7104
 \mathchardef\PLambda="7103
 \mathchardef\PSigma="7106
 \mathchardef\POmega="710A
 \mathchardef\PUpsilon="7107
                  
 \def\PB      {\ensuremath{B}\xspace}                 
                  
 \def\PD      {\ensuremath{D}\xspace}

 \def\PK      {\ensuremath{K}\xspace}

 \def\Pi      {\ensuremath{i}\xspace}

}

\makeatletter
\ifcase \@ptsize \relax% 10pt
  \newcommand{\miniscule}{\@setfontsize\miniscule{4}{5}}% \tiny: 5/6
\or% 11pt
  \newcommand{\miniscule}{\@setfontsize\miniscule{5}{6}}% \tiny: 6/7
\or% 12pt
  \newcommand{\miniscule}{\@setfontsize\miniscule{5}{6}}% \tiny: 6/7
\fi
\makeatother

\DeclareRobustCommand{\optbar}[1]{\shortstack{{\miniscule (\rule[.5ex]{1.25em}{.18mm})}
  \\ [-.7ex] $#1$}}

  \def\Kbar    {{\kern 0.2em\overline{\kern -0.2em \PK}{}}\xspace}

\def\KorKbar    {\kern 0.18em\optbar{\kern -0.18em K}{}\xspace}

  \def\Dbar    {{\kern 0.2em\overline{\kern -0.2em \PD}{}}\xspace}
\def\D       {{\ensuremath{\PD}}\xspace}

\def\DorDbar    {\kern 0.18em\optbar{\kern -0.18em D}{}\xspace}
\def\Dz      {{\ensuremath{\D^0}}\xspace}
\def\Dzb     {{\ensuremath{\Dbar{}^0}}\xspace}

\def\Bbar    {{\ensuremath{\kern 0.18em\overline{\kern -0.18em \PB}{}}}\xspace}

\def\BorBbar    {\kern 0.18em\optbar{\kern -0.18em B}{}\xspace}

  \def\Y#1S{\ensuremath{\PUpsilon{(#1S)}}\xspace}% no space before {...}!

\def\Lbar        {{\ensuremath{\kern 0.1em\overline{\kern -0.1em\PLambda}}}\xspace}
\def\LorLbar    {\kern 0.18em\optbar{\kern -0.18em \PLambda}{}\xspace}

\def\to                 {\ensuremath{\rightarrow}\xspace}

\def\AT#1     {\ensuremath{A_{\mathrm{T}}^{#1}}\xspace}           % 2

\def\C#1      {\ensuremath{\mathcal{C}_{#1}}\xspace}                       % 9
\def\Cp#1     {\ensuremath{\mathcal{C}_{#1}^{'}}\xspace}                    % 7
\def\Ceff#1   {\ensuremath{\mathcal{C}_{#1}^{\mathrm{(eff)}}}\xspace}        % 9  
\def\Cpeff#1  {\ensuremath{\mathcal{C}_{#1}^{'\mathrm{(eff)}}}\xspace}       % 7
\def\Ope#1    {\ensuremath{\mathcal{O}_{#1}}\xspace}                       % 2
\def\Opep#1   {\ensuremath{\mathcal{O}_{#1}^{'}}\xspace}                    % 7

\newcommand{\tev}{\ifthenelse{\boolean{inbibliography}}{\ensuremath{~T\kern -0.05em eV}\xspace}{\ensuremath{\mathrm{\,Te\kern -0.1em V}}}\xspace}
\newcommand{\gev}{\ensuremath{\mathrm{\,Ge\kern -0.1em V}}\xspace}
\newcommand{\mev}{\ensuremath{\mathrm{\,Me\kern -0.1em V}}\xspace}
\newcommand{\kev}{\ensuremath{\mathrm{\,ke\kern -0.1em V}}\xspace}
\newcommand{\ev}{\ensuremath{\mathrm{\,e\kern -0.1em V}}\xspace}
\newcommand{\gevc}{\ensuremath{{\mathrm{\,Ge\kern -0.1em V\!/}c}}\xspace}
\newcommand{\mevc}{\ensuremath{{\mathrm{\,Me\kern -0.1em V\!/}c}}\xspace}
\newcommand{\gevcc}{\ensuremath{{\mathrm{\,Ge\kern -0.1em V\!/}c^2}}\xspace}
\newcommand{\gevgevcccc}{\ensuremath{{\mathrm{\,Ge\kern -0.1em V^2\!/}c^4}}\xspace}
\newcommand{\mevcc}{\ensuremath{{\mathrm{\,Me\kern -0.1em V\!/}c^2}}\xspace}

\def\gsim{{~\raise.15em\hbox{$>$}\kern-.85em
          \lower.35em\hbox{$\sim$}~}\xspace}
\def\lsim{{~\raise.15em\hbox{$<$}\kern-.85em
          \lower.35em\hbox{$\sim$}~}\xspace}

\def\tell1  {TELL1\xspace}
\def\ukl1   {UKL1\xspace}

%% file: econfmacros.tex
\def\beq{\begin{equation}}
\def\eeq#1{\label{#1}\end{equation}}
\def\eeqn{\end{equation}}

\def\beqa{\begin{eqnarray}}
\def\eeqa#1{\label{#1}\end{eqnarray}}
\def\eeqan{\end{eqnarray}}

\let\bar=\overbar

\def\D{{\cal D}}

\def\Dslash{\not{\hbox{\kern-4pt $D$}}}
\def\dslash{\not{\hbox{\kern-2pt $\del$}}}

\def\msb{{\bar{\ssstyle M \kern -1pt S}}}